\documentclass[11pt]{article}
\usepackage{blois,epsfig}
\usepackage{graphicx}

\def\Journal#1#2#3#4{{#1} {\bf #2}, #3 (#4)}

\def\be{\begin{equation}}
\def\ee{\end{equation}}
\def\bea{\begin{eqnarray}}
\def\eea{\end{eqnarray}}

\begin{document}
\vspace*{4cm}
\title{Preparation to the CMB PLANCK data analysis, bias due to the
  galactic polarized emissions}

\author{ L. Fauvet, J.-F. Mac\'ias P\'erez, F.-X. D\'esert }

\address{LPSC, Universit\'e Joseph Fourier Grenoble 1, CNRS/IN2P3,Institut National Polytechnique de Grenoble, 53 avenue des Martyrs,\\
38026 Grenoble cedex, France}

\maketitle\abstracts{
The PLANCK satellite mission has been launched the 14th of May 2009 and
is dedicated to the measurement of the Cosmic Microwave Background
(CMB) in temperature and polarization. The presence of diffuse
galactic polarized emission contaminates the measurement of the CMB
anisotropies, in particular in polarization. Therefore a good
knowledge of these emissions is needed to the accuracy
required for PLANCK. In this context, we have developed and
implemented a coherent 3D model of the two main polarized galactic
emissions : synchrotron radiation and thermal dust. We have compared these
models to the WMAP and ARCHEOPS data and to the 408 MHz all-sky continuum
survey. From this, we are
able to estimate the contribution of polarized foreground emissions to
the polarized CMB radiation measured with PLANCK.}

\section{Introduction}

The PLANCK satellite which is currently in flight, should give the most
accurate measurement of the anisotropies of the CMB in temperature and
polarization with a sensitivity of $2 \mu K$ and an angular resolution
of 5 arcmin~\cite{bluebook}. It will cover a large range of frequencies between 30 and
857 GHz and it will allow us to study the Galactic
microwave emissions. In order to obtain the optimal
sensitivity it is necessary for us to estimate the foreground emissions and
the residual contamination due to these foreground emissions. Indeed
when, for the full sky, these emissions have the same order
of magnitude than the CMB in temperature, they dominate by a factor 10 in
polarization~\cite{bluebook}. The principal polarized Galactic microwave emissions come
from 2 effects : thermal dust emission and synchrotron emission. The synchrotron is well constrained by the 408 MHz all-sky continuum survey~\cite{haslam}, by Leider between 408 MHz and 1.4 GHz~\cite{wolleben}, by Parkes at 2.4 GHz~\cite{duncan1999}, by the MGLS {\it Medium Galactic Latitude Survey} at
1.4 GHz~\cite{uyaniker} and by the satellite WMAP {\it Wilkinson Microwave Anisotropies Probe} (see e.g.~\cite{hinshaw}). The synchrotron emission is due to ultrarelativistic
electrons spiraling in a large-scale magnetic field. The thermal dust
emission which has already been well constrained by IRAS~\cite{schlegel}, COBE-FIRAS~\cite{boulanger} and
ARCHEOPS~\cite{benoit}, is due to dust grains
which interact with the Galactic magnetic field and emit a polarized
submillimetric radiation~\cite{boulanger}. The polarization of these two
types of radiation is orthogonal to the field lines. To obtain a realistic model of
these emissions we must build a model based on a 3D modelisation of the
Galactic magnetic field and the matter density in the
Galaxy. The models are optimized with respect to preexisting data and
can then be used to estimate the bias due to
these emissions on the CMB measurement.

\section{3D modelling of the Galaxy}
\label{sec:model}

\indent A polarized emission can be described by the Stokes
parameters I,Q and U ~\cite{kosowsky}. For the polarized foreground
emissions integrated along the line of sight we obtain, for synchrotron:
\begin{eqnarray}
\label{eq:map_sync}
\centering
 I_s &=& I_{\mathrm{Has}} \left(\frac{\nu_s}{0,408}\right)^{\beta_s},\\
Q_s &=& I_{\mathrm{Has}}
\left(\frac{\nu_s}{0,408}\right)^{\beta_s}\frac{\int \cos(2\gamma)p_s n_e\left(B_l^2 + B_t^2 \right)}{\int n_e\left(B_l^2 + B_t^2 \right)} ,\\
U_s &=& I_{\mathrm{Has}}
\left(\frac{\nu_s}{353}\right)^{\beta_s}\frac{\int \sin(2\gamma)p_s n_e\left(B_l^2 + B_t^2 \right)}{\int n_e\left(B_l^2 + B_t^2 \right)},
\end{eqnarray}

\noindent where $B_n$, $B_l$ and  $B_t$ are the magnetic field components
along, longitudinal and transverse to the ligne of sight. $p_s$ is the
polarization fraction set to 75\%. $I_{Has}$ is the template
map~\cite{haslam}. The maps are extrapolated at all the Planck
frequencies using the spectral index $\beta_s$ which is a free parameter of the model.

For the thermal dust emission : \\
\begin{eqnarray}
\centering
 I_d &=& I_{sfd} \left( \frac{\nu_d}{353} \right)^{\beta_d},\\
Q_d &=& I_{sfd} \left( \frac{\nu_d}{353}\right)^{\beta_d} \int n_d\frac{\cos(2 \gamma) \sin^2(\alpha) f_{\mathrm{norm}}p_d}{n_d} ,\\
U_d &=& I_{sfd} \left(\frac{\nu_d}{353}\right)^{\beta_d}  \int n_d \frac{\sin(2 \gamma) \sin^2(\alpha) f_{\mathrm{norm}}p_d}{ \int n_d} ,
\end{eqnarray}

\noindent where the polarization fraction $p_d$ is set to 10 \%,
$\beta_d$ is the spectral index (set at 2.0) and $f_{norm}$ is an
empiric factor, fit to the ARCHEOPS data. The
$I_{sfd}$ map is the model 8 of~\cite{schlegel}.\\ 

\indent The models are based on an exponential distribution of
relativistic electrons on the Galactic disk,
following~\cite{drimmel}, where the radial scale $h_r$ is a free
parameter. The distribution of dust grains $n_d$ is
also exponential~\cite{benoit}. The Galactic magnetic field is composed of two
parts: a regular component and a turbulent component. The regular
component is based on the WMAP team model~\cite{page} which is close to a
logarithmic spiral to reproduce the shape of the spiral
arms~\cite{han2006}. The pitch angle $p$ between two spiral arms is a free
parameter of the model. The turbulent component is described by
a law of Kolmogorov~\cite{han2006} spectrum of relative amplitude $A_{turb}$.

\section{Comparison to data}
\label{sec:test}

\begin{figure}[htbf]
\centering
\includegraphics[height=11cm,width=9cm]{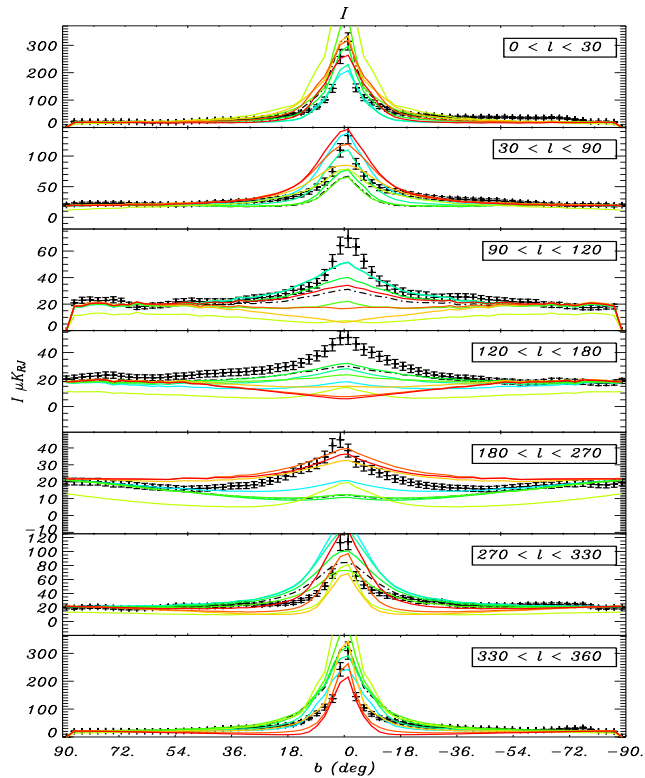}\caption{Galactic profiles in temperature at 408 MHz Haslam data in black and our synchrotron emission moedl for various values of the pitch angle $p$ {\it (form green to red)}.\label{fig:gal_has}}
\end{figure}

\indent We computed Galactic profiles in temperature and polarization
for various bands in longitude and latitude and various values of the free
parameters. In order to optimize these 3D models we compare them to
Galactic profiles computed using preexisting data using a $\chi^2$
test. For the synchrotron emission in temperature, we use the 408 MHz all-sky
continuum survey~\cite{haslam} as shown on
Figure~\ref{fig:gal_has}. In polarization we compared to the K-band
WMAP 5 years data. The thermal dust emission model is optimized using the
polarized ARCHEOPS data~\cite{benoit} at 353 GHz.

\indent The best fit parameters for the 3D model in polarization
are given in Table~\ref{tab:param}. The results are coherent for
the 3 sets of data. In particular we obtain compatible results
for the synchrotron and thermal dust emission models. $A_{turb}$ and
$h_r$ are badly constrained as was already the case in Sun {\it et al}~\cite{sun}. The
best fit value of the pitch angle $p$ is compatible with results
obtained by other studies~\cite{sun,page}. 
The best fit value for the spectral index of the synchrotron
emission is lower than the value found by ~\cite{sun,page}, but this is
probably due to the choice of normalisation for the regular componenent
of the magnetic field. With these models we reproduce the global structure of the
data (see for instance the Figure~\ref{fig:gal_has}) appart from the
Galactic center.

\begin{table*}[h]
\begin{center}
\caption{Best fit parameters for synchrotron and thermal dust emission models.\label{tab:param}}
\vspace{0.3cm}

\begin{tabular}{|c|c|c|c|c|c|} \hline
$$  & $ p (deg)$& $A_{turb} $  & $h_r$  &    $\beta_s$  & $\chi^2_{min}$  \\\hline
$WMAP$ & $ -30.0^{+40.0}_{-30.0}$ & $< 1.25$ (95.4 \% CL) & $ >1$ (95.4 \%
CL) &  $-3.4^{+0.1}_{-0.8}$ & $5.72$     \\\hline
$HASLAM$ & $ -10.0^{+70.0}_{-60.0}$   & $< 1.25$ (95.4 \% CL) &  $ 5.0^{+15.0}_{-2.0} $ & $\emptyset$ & $5.81$ \\\hline
$ARCHEOPS$ & $ -20^{+80}_{-50}$   & $ < 2.25 (95.4 \% CL)$ & $\emptyset$ & $\emptyset$ &  $ 1.98$          \\\hline

\end{tabular}
\end{center}
\end{table*}

\section{Conclusions}

\begin{figure}[htbf]
\centering
\includegraphics[height=9cm,width=13cm]{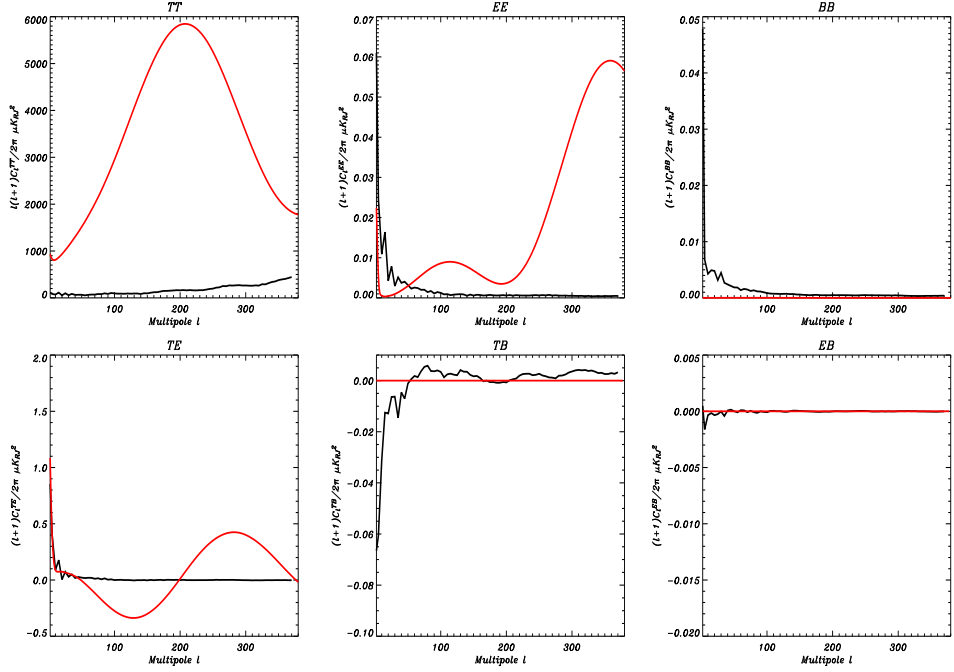}
\caption{Clockwise from top left : power spectra
  $C^{TT}_l$,$C^{EE}_l$,$C^{BB}_l$,$C^{TE}_l$,$C^{TB}_l$,$C^{EB}_l$ at
  143 GHz for $|b|<15 deg$ (see text for details).\label{fig:spect_cmb_for}}
\end{figure}

\indent From the above best fit parameters we estimate the contamination of the CMB PLANCK data by the
polarized galactic emissions. We compared power spectra computed with
 simulations of the CMB PLANCK data \footnote{We used cosmological
  parameters for a model $\Lambda$CDM like proposed in ~\cite{komatsu}
  with a ratio tensor-scalar of 0.03.}. Figure~\ref{fig:spect_cmb_for}
shows the temperature and polarization power spectra at 143 GHz for
the CMB simulation (\emph{red})
and the Galactic foreground emissions applying a Galactic cut $|b|<15
deg$ (\emph{black}). The foreground contamination seems to be weak but for the BB
modes an accurate foreground substraction is extremely important for
the detection of the primordial gravitational waves. More details can
be found in~\cite{fauvet}.

\end{document}